\documentclass[aps,prb,reprint,superscriptaddress,showpacs,amsfonts,amsmath,amssymb]{revtex4-1}

\usepackage{graphicx}

\begin{document}


\title{Enhancement of critical current density and vortex activation energy in proton-irradiated Co-doped BaFe$_{2}$As$_2$}


\author{Toshihiro Taen}
\affiliation{Department of Applied Physics, The University of Tokyo, 7-3-1 Hongo, Bunkyo-ku, Tokyo 113-8656, Japan}
\author{Yasuyuki Nakajima}
\affiliation{Department of Applied Physics, The University of Tokyo, 7-3-1 Hongo, Bunkyo-ku, Tokyo 113-8656, Japan}
\affiliation{JST, Transformative Research-Project on Iron Pnictides (TRIP), 7-3-1 Hongo, Bunkyo-ku, Tokyo 113-8656, Japan}
\author{Tsuyoshi Tamegai}
\affiliation{Department of Applied Physics, The University of Tokyo, 7-3-1 Hongo, Bunkyo-ku, Tokyo 113-8656, Japan}
\affiliation{JST, Transformative Research-Project on Iron Pnictides (TRIP), 7-3-1 Hongo, Bunkyo-ku, Tokyo 113-8656, Japan}
\author{Hisashi Kitamura}
\affiliation{Radiation Measurement Research Section, National Institute of Radiological Sciences, 4-9-1, Anagawa, Inage-ku, Chiba 263-8555, Japan}



\date{\today}

\begin{abstract}
The effect of proton irradiation in Ba(Fe$_{0.93}$Co$_{0.07}$)$_{2}$As$_{2}$ single crystals is reported.
We analyze temperature dependence of current density and normalized flux relaxation rate in the framework of collective creep model.
Glassy exponent and barrier height for flux creep are directly determined by Maley's method.
Our model functions for barrier height and critical current density in the absence of flux creep are explained by the superposition of $\delta T_c$- and $\delta l$-pinning.
We also approach true critical current density by means of generalized inversion scheme,
and the obtained result is in reasonable agreement with our model function.
Proton irradiation effect on temperature dependence of current density and normalized relaxation rate can be summarized as doubling of barrier height at the beginning of flux creep.
\end{abstract}


\maketitle

In high-temperature superconductors, many interesting phenomena in vortex dynamics are discovered such as giant-flux creep, thermally activated flux flow,
and the theories to describe them have been elaborated in the last decades.~\cite{RevModPhys.66.1125}
Especially, collective pinning with weak pinning potential by the quenched disorder and collective creep of vortex bundles give rise to intriguing experimental results, such as "plateau" observed in temperature dependent normalized relaxation rate ($S\equiv |\mathrm{d}\ln M/\mathrm{d}\ln t|$),~\cite{PhysRevB.42.6784}
in contrast to linear increase with temperature predicted by Anderson-Kim model in low-temperature superconductors.
Recently discovered iron-based superconductors (IBSs) have relatively high critical temperature ($T_c$) and large critical current density ($J_c$).
Besides, magnetization hysteresis loop in this system is quite similar to that in Y-Ba-Cu-O,
and magnetic relaxation measurements have revealed that IBSs also show giant-flux creep,
which implies that IBSs and cuprate superconductors share common vortex physics.
Moreover, how introduction of artificial pinning center affects flux dynamics or $J_c$ is also interesting.~\cite{SupercondSciTechnol10_A11,SupercondSciTechnol.25.084008}
In Y-Ba-Cu-O, $J_c$ is enhanced and glassy behavior remains basically the same after proton (H$^{+}$) irradiation,
which is known to introduce point defects.
This is also expected in IBSs.
In fact, Haberkorn {\it{et al.}} recently reported that H$^{+}$-irradiation does not affect to $H-T/T_c$ phase diagram in Ba(Fe$_{0.925}$Co$_{0.075}$)$_2$As$_2$.~\cite{PhysRevB.85.014522}
It is important to clarify how vortex dynamics is affected by H$^+$-irradiation in IBSs.

As we mentioned above, IBSs are suitable candidates to check whether glassy behavior of vortices is universal in all high-temperature superconductors.
Since this system is twin free and less anisotropic,
it enables us to discuss intrinsic pinning and dynamic properties of vortices without complication.
However, it is difficult to synthesize large and clean crystals especially in $Ln$FeAsO ($Ln$ is lanthanoid, so-called '1111' system).~\cite{Kamihara2008ils}
This prevents us from discussing vortex dynamics due to strong inhomogeneities unless we use local probes.
Since high-quality single crystals are readily available in so-called '122' crystals $AE$Fe$_2$As$_2$ ($AE$ : alkaline earths),~\cite{PhysRevLett.101.107006,PhysRevLett.101.117004} 
it is possible to discuss details of vortex pinning also in IBSs with global magnetic measurements.
Actually, homogeneous flow of superconducting current in this system has been confirmed by magneto-optical measurement.~\cite{JPSJ.78.023702}
This is why we choose optimally Co-doped BaFe$_2$As$_2$ single crystal.

In this paper, we report the effect of proton (H$^+$) irradiation in Ba(Fe$_{0.93}$Co$_{0.07}$)$_{2}$As$_{2}$ single crystals.
We analyze it in the framework of collective creep theory with temperature dependent shielding current $J$ and $S$.
Glassy exponent and barrier height for flux creep are directly determined.
Our model functions for barrier height and critical current density in the absence of flux creep are explained by the superposition of $\delta T_c$- and $\delta l$-pinning.
We also approach the true critical current density by means of generalized inversion scheme,
and the obtained result is in reasonable agreement with our model function.
Effects of H$^+$ irradiation on $J(T)$ and $S(T)$ can be summarized as doubling of barrier height at the beginning of flux creep.

Optimally Co-doped BaFe$_2$As$_2$ single crystals were grown by FeAs/CoAs self-flux method.
Fundamental properties of this system have been reported elsewhere.~\cite{JPSJ.78.023702,Nakajima2010S408}
All samples are cleaved to be thin plates with thickness less than $\sim 20 \;\mu$m.
This value is much smaller than the projected range of 3 MeV H$^{+}$ for Ba(Fe$_{0.93}$Co$_{0.07}$)$_{2}$As$_{2}$ of $\sim 50 \;\mu$m,
calculated by the stopping and range of ions in matter-2008.~\cite{SRIM}
3 MeV H$^{+}$ irradiation was performed parallel to $c$-axis at 40 K at NIRS-HIMAC.
The total dose of the measured sample is $1.2 \times 10^{16}$ cm$^{-2}$.~\cite{PhysRevB.82.220504}
Magnetization is measured in a commercial SQUID magnetometer (MPMS-XL5, Quantum Design) with a magnetic field parallel to $c$-axis.
Main features of vortex dynamics in H$^{+}$-irradiated sample analyzed in this work have been briefly reported in Ref. {\onlinecite{PhysicaC.471.784}}.
To clarify the effect of H$^{+}$ irradiation on the vortex system,
we also measured a pristine sample similar to the "unirradiated" sample in Ref. {\onlinecite{PhysRevB.80.012510}}.
Current densities calculated by the Bean model are denoted as $J_s$ in field-sweep measurements (s is an abbreviation for sweep), or simply $J$ in relaxation measurements.

\begin{figure}
\includegraphics[width=7.5cm]{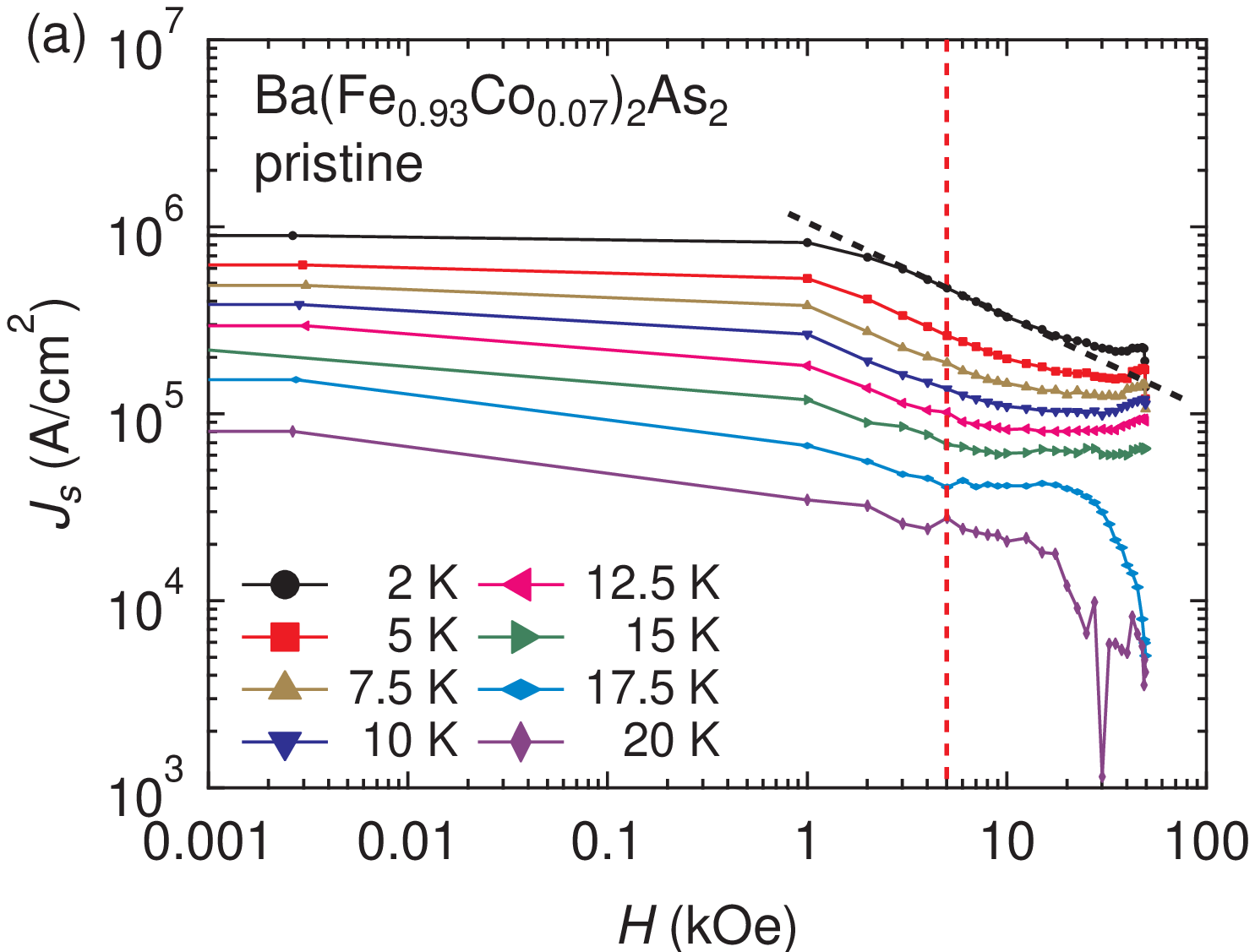}\\%
\includegraphics[width=7.5cm]{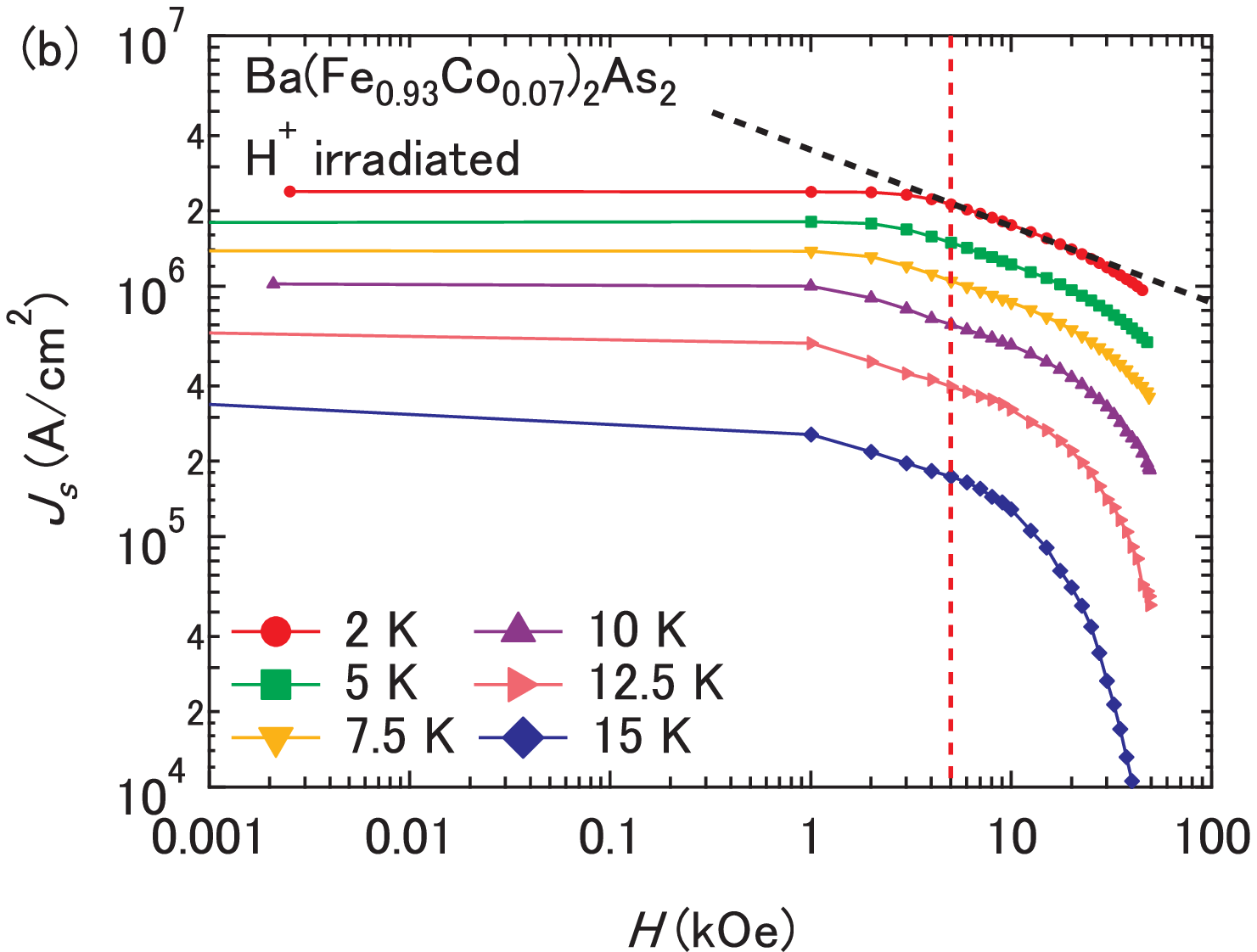}%
\caption{\label{Figure1}Field dependence of $J_s$ in (a)pristine and (b)H$^{+}$ irradiated Ba(Fe$_{0.93}$Co$_{0.07}$)$_{2}$As$_{2}$ at several temperatures. Dotted line on $\bullet$ (2 K) shows power-law decay of (a) $H^{-0.5}$ and (b) $H^{-0.3}$, respectively. Vertical line indicates the field where we discuss the vortex dynamics, $H = 5$ kOe.}
\end{figure}

Figure \ref{Figure1} shows magnetic field dependence of $J_s$ in (a) pristine and (b) H$^{+}$-irradiated Ba(Fe$_{0.93}$Co$_{0.07}$)$_{2}$As$_{2}$.
It is obvious that H$^+$ irradiation enhances $J_s$ from $1\times 10^6$ A/cm$^2$ to $2.5\times 10^6$ A/cm$^2$ at 2 K under zero-field.
In pristine sample, $J_s$ is nearly constant below 1 kOe, followed by power law decay $H^{-\alpha}$ at a field of 2-10 kOe with $\alpha \sim 0.5$.
As discussed by van der Beek {\it{et al.}}, these behaviors at low-fields are attributed to sparse strong-point pinning centers, as in the case of Y-Ba-Cu-O films.~\cite{PhysRevB.81.174517,PhysRevB.66.024523}
At a glance, it seems inappropriate to analyze it in the framework of collective-pinning-collective-creep and vortex glass theory.
However, since strong-point-pinning contribution for temperature ($T$) dependence of current density ($J$) is smaller than weak-collective pinning (see Fig.~6 or 9 in Ref.~\onlinecite{PhysRevB.81.174517}),
we can approximate $J(T)$ only by the contribution from collective creep (pinning).
In H$^+$-irradiated sample, it is basically the same as the pristine one, although there is a wide-crossover region with $\alpha \sim 0.3$ between low-field plateau and $H^{-0.5}$ region.
Such a weak field dependence has been also observed in YBa$_2$Cu$_3$O$_7$ and YBa$_2$Cu$_4$O$_8$ films by Griessen {\it{et al.}}, and they concluded that single vortex creep is achieved in this region.~\cite{PhysRevLett.72.1910}


\begin{figure}
\includegraphics[width=7.5cm]{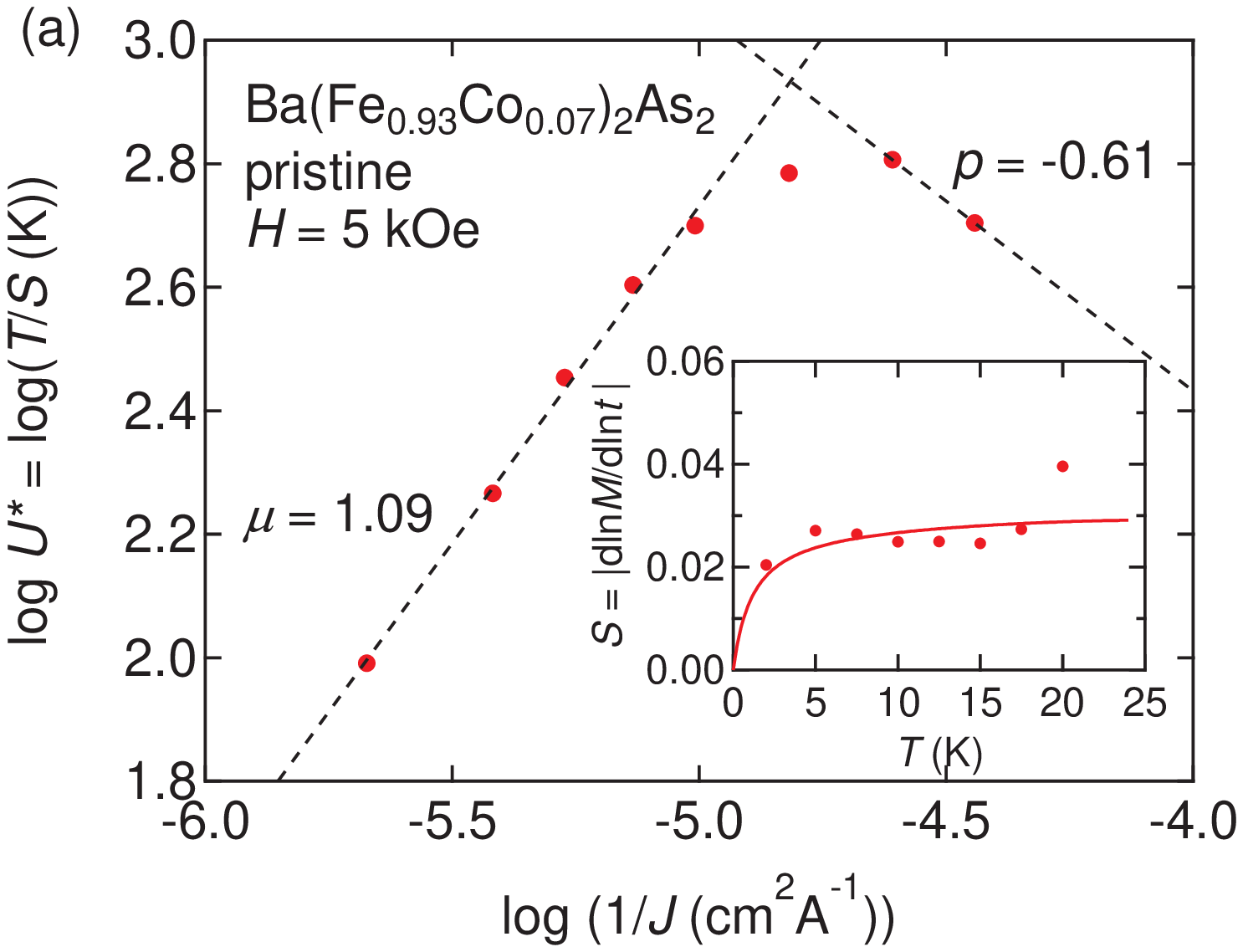}\\%
\includegraphics[width=7.5cm]{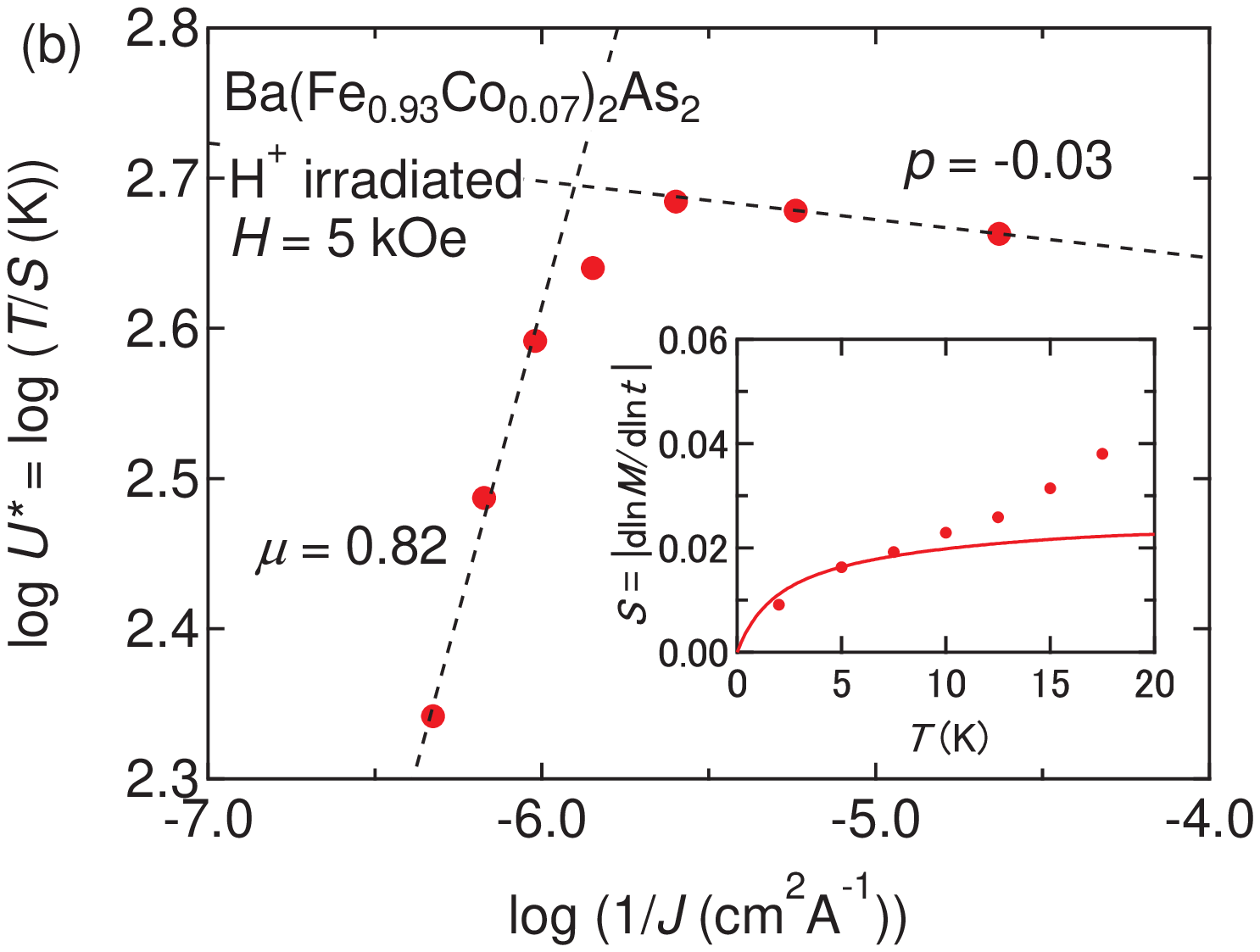}%
\caption{\label{Figure2}Inverse current-density dependence of effective pinning energy $U^*$ in (a)pristine and (b)H$^{+}$ irradiated Ba(Fe$_{0.93}$Co$_{0.07}$)$_{2}$As$_{2}$. Inset: Temperature dependence of normalized relaxation rate $S$. Solid line indicates fitting by Eq.~\eqref{glassS}.}
\end{figure}

To elucidate the vortex dynamics, it is important to measure (static) magnetic relaxation rate $S \equiv |\textrm{dln}M/\textrm{dln}t|$ in both samples,
where  $M$ is magnetization, $t$ is time from the moment when the critical state is prepared.
In order to discuss temperature dependence of vortex dynamics, we have to fix magnetic field.
However, as we mentioned above, there is a strong pinning background,
so that we have to select a field where field dependence of $J_s$ is similar for all temperatures to exclude field-dependent strong-point-pinning effect.
Besides, we should carefully keep away from fish-tail effect with non-monotonic field dependence of $J_s$ at high fields and self-field effect at low fields, which disturb direct extraction of typical parameters for vortex dynamics.
Based on these consideration, we select $H = 5$ kOe in both samples, shown as vertical broken lines in Fig.~\ref{Figure1}.
Insets of Fig.~\ref{Figure2} show temperature dependence of $S$.
According to collective pinning theory,~\cite{RevModPhys.68.911} 
this is described as
\begin{equation}
	S = \frac{T}{U_0 +\mu T \ln (t/t_{\textrm{eff}})},
	\label{glassS}
\end{equation}
where $U_0$ is temperature-dependent flux activation energy in the absence of flux creep,
$\mu > 0$ is a glassy exponent for elastic creep, and $t_{\textrm{eff}}$ is effective hopping attempt time.
One of the most remarkable results extracted from this equation is the prediction of plateau in the intermediate temperature range if $U_0\ll T$.~\cite{PhysRevB.42.6784}
The value of plateau $S \sim 1/(\mu \ln(t/t_{\textrm{eff}}))$ falls in the range of 0.02-0.04 theoretically, which has been confirmed in Y-Ba-Cu-O.~\cite{PhysRevB.42.6784}
The inset of Fig.~\ref{Figure2}(a) is consistent with this behavior quantitatively, as observed in other IBSs.~\cite{PhysRevB.80.012510,PhysRevB.78.224506,PhysRevB.84.094522}
This proves the validity of applying collective pinning theory to IBSs.
Here we emphasize that it is quite important to determine the value of $\mu$ in discussing vortex dynamics,
since $\mu$ includes information on the size of vortex bundle in collective pinning theory.
In three-dimensional system, it is predicted as $\mu = 1/7, 3/2, 7/9$ for single-vortex, small-bundle, and large-bundle regime, respectively.~\cite{PhysRevLett.63.2303}
Inverse current-density dependence of effective pinning energy $U^{*} = T/S$ is convenient to extract this value.

We can define inverse power-law form of flux activation energy $U(J)$ as
\begin{equation}
	U(J) = \frac{U_0}{\mu}\left[(J_{c0}/J)^{\mu}-1\right].
	\label{nonlinearU(J)}
\end{equation}
Combining this with $U = T \ln (t/t_{\textrm{eff}})$ extracted from Arrhenius relation,
we can deduce the so-called "interpolation formula":
\begin{equation}
	J(T,t) = \frac{J_{c0}}{\left[1+(\mu T/U_0)\ln (t/t_{\textrm{eff}})\right]^{1/\mu}},
	\label{interpolationformula}
\end{equation}
where $J_{c0}$ is temperature dependent critical current density in the absence of flux creep.
From Eqs.~\eqref{nonlinearU(J)} and \eqref{interpolationformula},
\begin{equation}
	U^{*} = U_0+\mu T \ln(t/t_{\textrm{eff}}) = U_0\left(J_{c0}/J\right)^{\mu}
\end{equation}
is derived.
Thus the slope in double logarithmic plot of $U^*$ vs. $1/J$ gives the value of $\mu$, shown in the main panels of Fig.~\ref{Figure2}.
In this way, we evaluate $\mu = 1.09$ and 0.82 for pristine and H$^{+}$-irradiated samples, respectively.
Note that $\mu \simeq 1$ in pristine crystal is often reported in Y-Ba-Cu-O~\cite{PhysRevB.47.14440} and IBSs~\cite{PhysRevB.82.054513}.
Contrary to the above prediction of $\mu > 0$, negative slope is observed at small $J$.
This negative slope is often denoted as $p$ in plastic creep theory with $p = -0.5$,
and confirmed experimentally.~\cite{PhysRevLett.77.1596}
Our evaluation of $p = -0.61$ in pristine sample is very similar to this value.

\begin{figure}
\includegraphics[width=7.5cm]{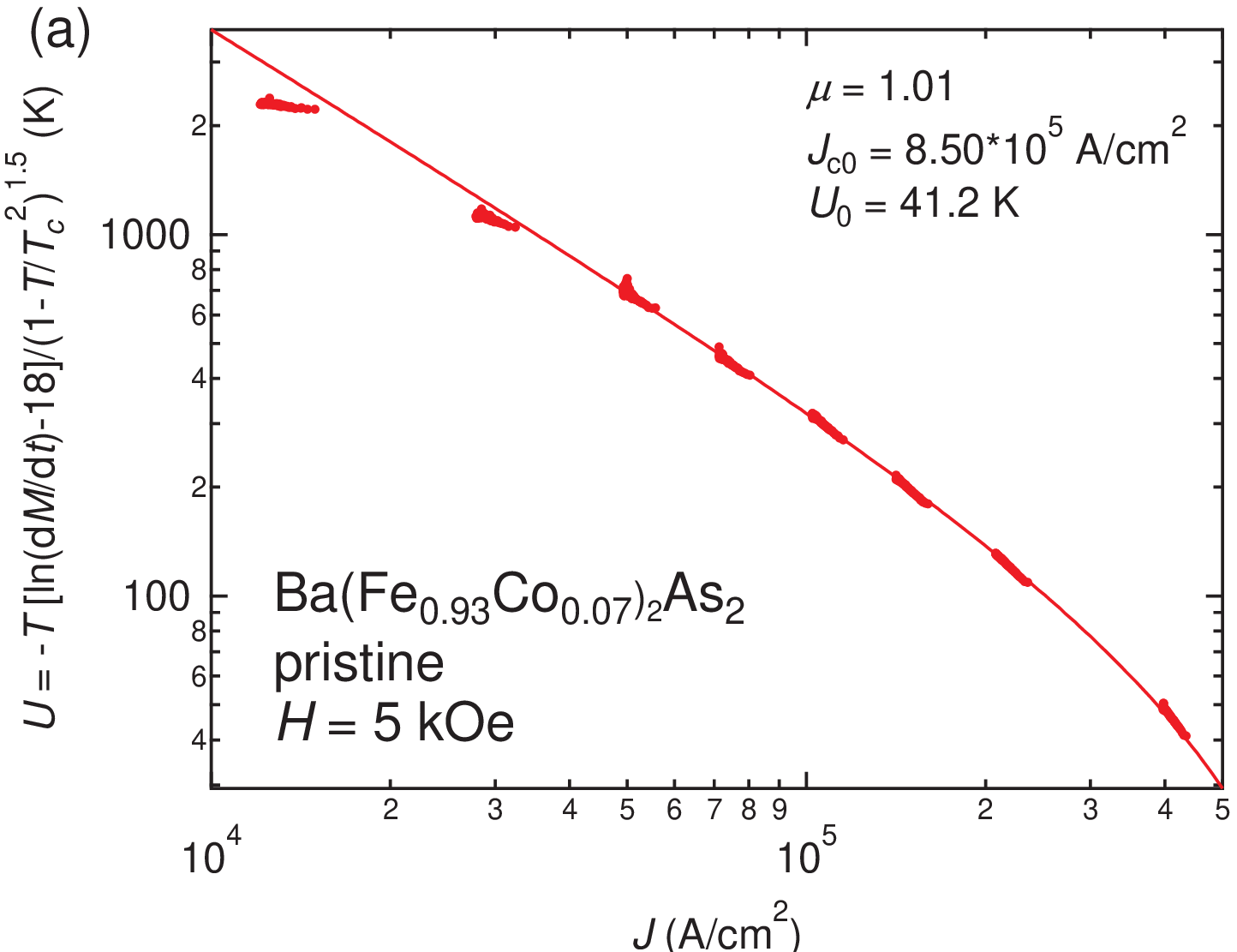}\\%
\includegraphics[width=7.5cm]{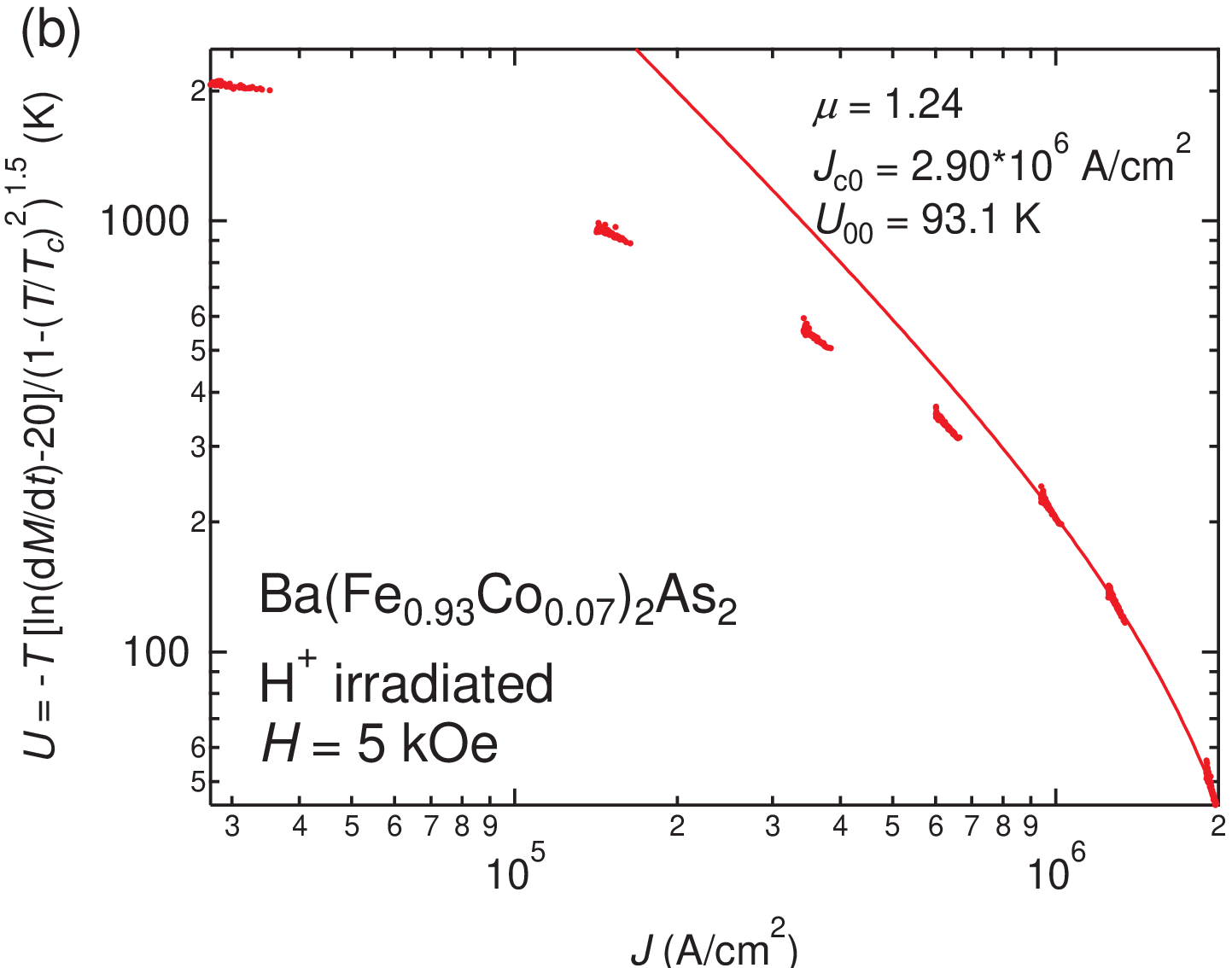}%
\caption{\label{Figure3}Current density dependence of $U$ in (a) pristine and (b) H$^{+}$-irradiated Ba(Fe$_{0.93}$Co$_{0.07}$)$_{2}$As$_{2}$ constructed by extended Maley's method. Solid line indicates power-law fitting in large $J$ region.}
\end{figure}

To determine actual flux activation energy, we employ extended Maley's method.~\cite{SupercondSciTechnol.23.025033}
Since temperature dependence of $U$ is not considered in the original Maley's method,~\cite{PhysRevB.42.2639}
it is impossible to scale $U$ in a wide range of $J$ even if glassy exponent is unique.
In order to solve this problem, appropriate temperature dependences of $U_0$ and $J_{c0}$ are assumed as follows.
\begin{align}
	U_0(T) &= U_{00}[1-(T/T_c)^2]^n,
	\label{U0T}\\
	J_{c0}(T) &= J_{c00}[1-(T/T_c)^2]^n.
	\label{Jc0T}
\end{align}
In order to simplify the problem, we choose the same exponents in Eq.~\eqref{U0T} and Eq.~\eqref{Jc0T}. 
Here exponent $n$ is set to 3/2,
as in the case of Ref. \onlinecite{PhysRevB.47.14440},\onlinecite{SupercondSciTechnol.23.025033}, 
while $(1-T/T_c)^{3/2}$ is selected in Ba$_{1-x}$K$_{x}$Fe$_2$As$_2$.~\cite{PhysRevB.82.054513}
$U=-T\ln[\textrm{d}M(t)/\textrm{d}t]+CT$ and $C=\ln(B\omega a/2\pi r)$ is assumed as a constant, where $B$ is the magnetic induction, $\omega$ is the attempt frequency for vortex hopping, $a$ is the hopping distance, and $r$ is
the sample radius.
We select $C=18$ and 20 for pristine and H$^{+}$-irradiated samples, respectively.
Figures \ref{Figure3} show current density dependence of $U$ in (a)pristine and (b)H$^{+}$-irradiated Ba(Fe$_{0.93}$Co$_{0.07}$)$_{2}$As$_{2}$, respectively, constructed by extended Maley's method. 
Solid lines indicate power-law fitting to large $J$ region where the slope in Fig.~\ref{Figure2} is positive.
Note that deviation of the data from the fitting in the small $J$ region is reasonable since creep is plastic there.
The obtained glassy exponents are $\mu = 1.01$ and 1.24 for pristine and H$^{+}$-irradiated samples, respectively.
For pristine sample, this value is nearly the same as that obtained in Fig.~\ref{Figure2}(a), $\mu = 1.09$.
On the other hand, the change of $\mu$ by H$^{+}$ irradiation has an opposite trend.
Namely, the value of $\mu$ decreases in Fig.~\ref{Figure2}(b), while grows in Fig.~\ref{Figure3}(b) after H$^{+}$-irradiation.
This is because the vortex system in H$^{+}$-irradiated sample crossovers from elastic to plastic creep more gradually, as we can see in the main panel and inset of Fig.~\ref{Figure2}(b).
Hence we may underestimate $\mu$ and overestimate $p$ with the scheme of Fig.~\ref{Figure2}, and it is more reliable to estimate it from the extended Maley's method of Fig.~\ref{Figure3}, which uses more data points.
For this reason, we conclude that $\mu$ is slightly increased by H$^{+}$ irradiation. 
Additionally, slight increase of $\mu$ is consistent with the regime of measurement,
where it is closer to small bundle regime with $\mu = 3/2$, as we discussed in Fig.~\ref{Figure1}.
Other resultant parameters are summarized in Table \ref{Table1}.
With these $U_{00}$, temperature dependence of $S$ is fitted by Eq.~\eqref{glassS} with a single free parameter of plateau value $S^{\rm{sat}} = 1/\mu \ln(t/t_{\textrm{eff}})$ as shown in the inset of Fig. ~\ref{Figure2}.
The inverse of this value is also shown in Table \ref{Table1}.

\begin{table}
\caption{\label{Table1}Parameters obtained from extended Maley's method and inverse of plateau value in $S(T)$.}
\begin{ruledtabular}
\begin{tabular}{ccccc}
	Sample & $J_{c0}$ & $U_{00}$ & $\mu$ & $\mu \ln(t/t_{\textrm{eff}})$ \\
	 & (MA/cm$^2$) & (K) & & \\
	Pristine & 0.85 & 41.2 & 1.01 & 35 \\
	H$^{+}$ irradiated & 2.90 & 93.1 & 1.24 & 43 \\
\end{tabular}
\end{ruledtabular}
\end{table}

\begin{figure}
\includegraphics[width=7.5cm]{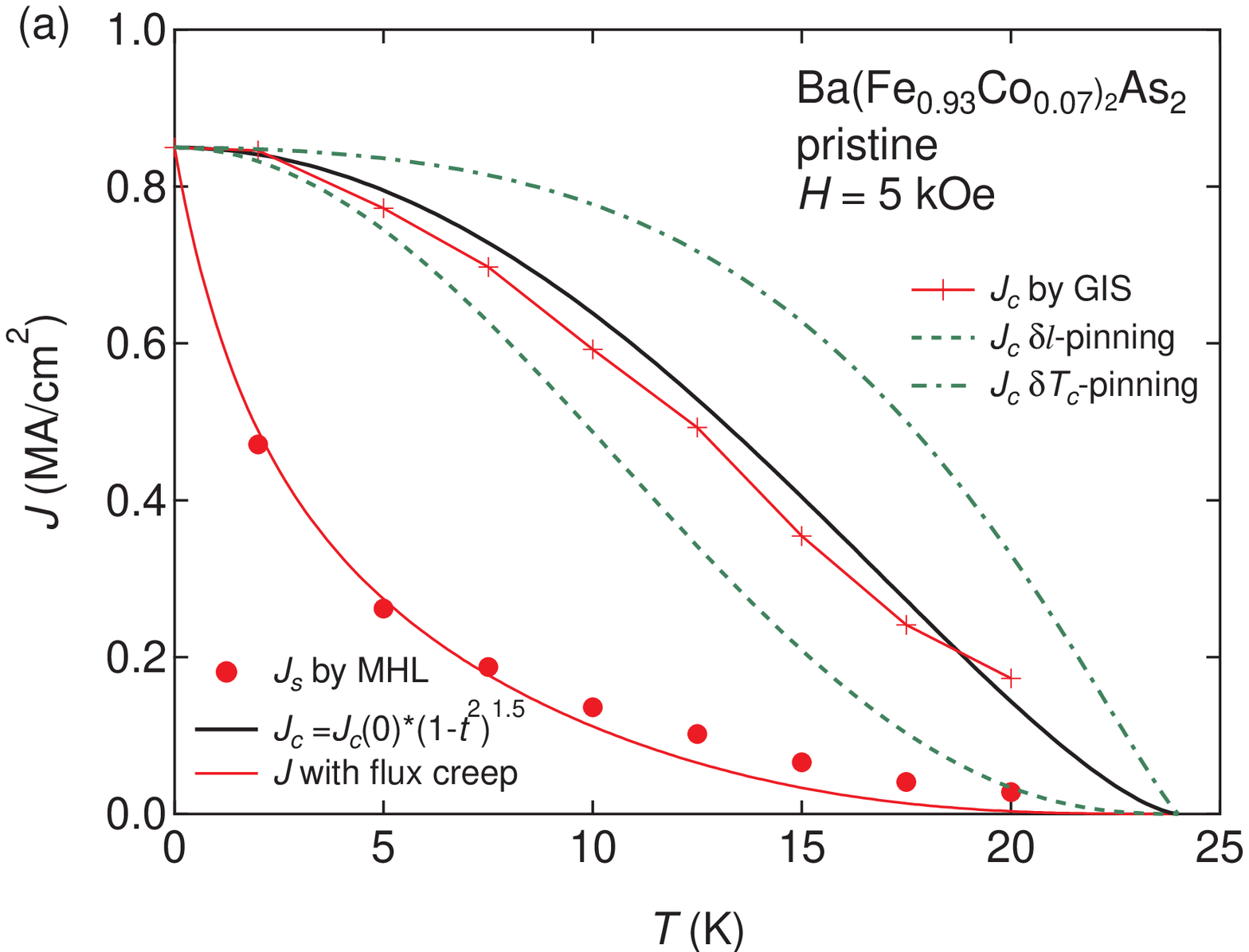}\\%
\includegraphics[width=7.5cm]{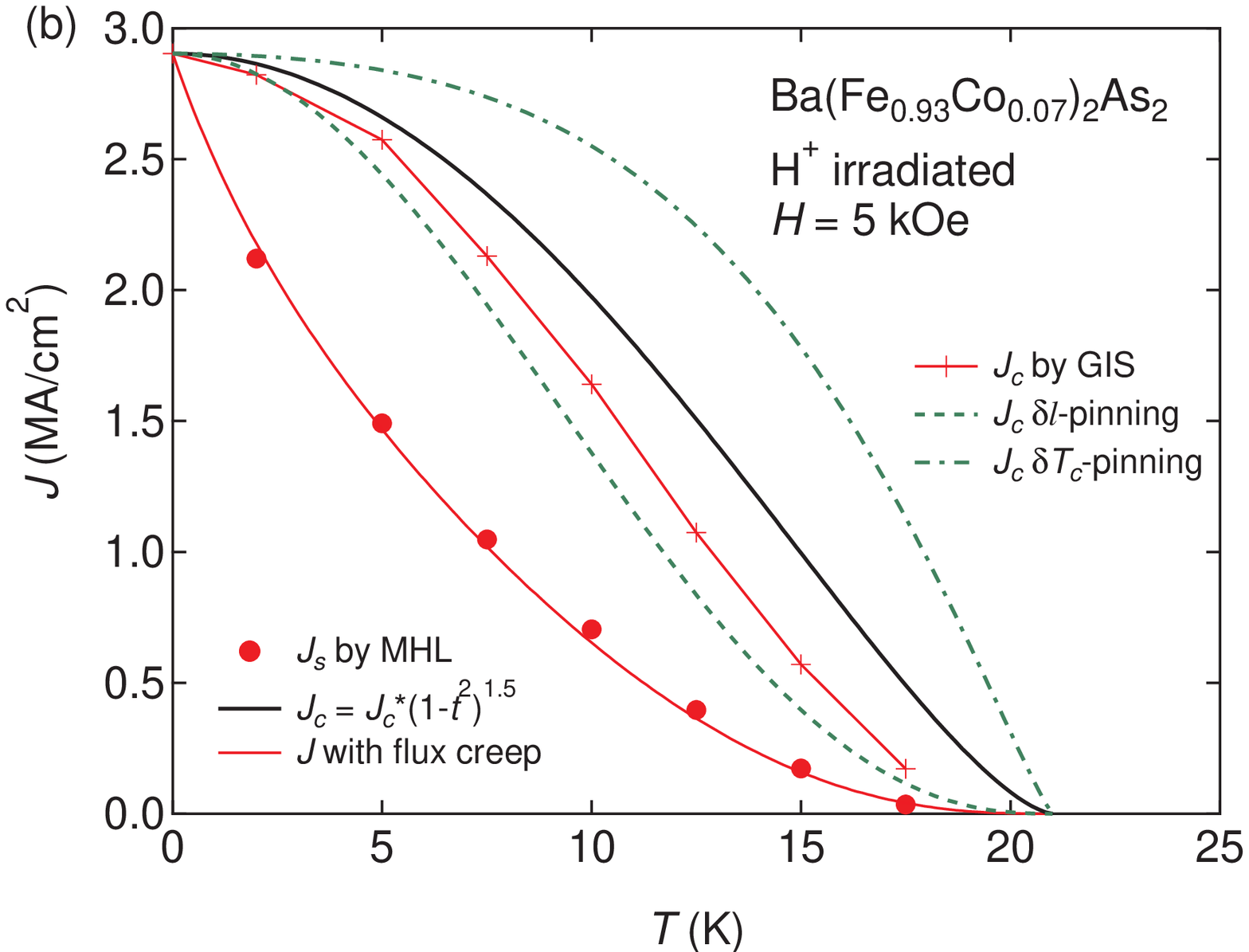}%
\caption{\label{Figure4}Temperature dependence of $J_s$ (\textbullet), model function of $J_c$ before and after creep with the parameters in Table \ref{Table1} (solid line), $\delta T_c$- and $\delta l$-pinning (dashed line) and $J_c$ reconstructed by GIS (+) in (a)pristine and (b)H$^{+}$-irradiated Ba(Fe$_{0.93}$Co$_{0.07}$)$_{2}$As$_{2}$.}
\end{figure}

Using parameters obtained above, we calculate $J$ after creep from attempt function of (true) critical current density Eq.~\eqref{Jc0T},
which is shown as lower solid line in Fig.~\ref{Figure4}.
In both cases, $J$ is reasonably reproduced, especially at high $J$ region (i.e. at low temperature).
This means that the present collective pinning/creep analysis is appropriate.
To get more insight into pinning mechanism in IBSs, we also show a function of $\delta T_c$- and $\delta l$-pinning in Fig.~\ref{Figure4}.
These functions are written as $J_c(t)/J_c(0) = (1-t^2)^{7/6}(1+t^2)^{5/6}$ and $(1-t^2)^{5/2}(1+t^2)^{-1/2}$, respectively.~\cite{PhysRevLett.72.1910}
From this figure, our model function of $J_c$ can be considered as a superposition of the two pinning mechanisms.
To discuss such a mechanism, generalized inversion scheme (GIS) is utilized.~\cite{PhysRevB.48.13178,PhysicaC.241.353}
Although in this scheme, we have to assume empirical temperature dependence of penetration depth $\lambda$ and coherence length $\xi$ as $\propto (1-t^4)^{-1/2}$ and $\propto (1+t^2)^{1/2}(1-t^2)^{-1/2}$, respectively,
we can directly reconstruct true critical current density $J_c$ from $J_s$ and discuss pinning mechanism.
When we assume $\ln (t/t_{\textrm{eff}}) \sim 23$ in the measurement with field sweeping,
and choose parameters for three-dimensional single vortex pinning,
$J_c$ is reconstructed as shown in Fig.~\ref{Figure4},
which are in reasonable agreement with the model function.
Similar analyses of pinning mechanism using GIS in pristine Ba(Fe$_{1-x}$Co$_x$)$_2$As$_2$ have been attempted in Ref. \onlinecite{PhysRevB.81.014503}.
They also conclude that both $\delta T_c$- and $\delta l$-mechanisms are working in this system.
Here, we want to compare our work with similar work by Haberkorn \textit{et al.}.~\cite{PhysRevB.85.014522}
In their work, temperature dependence of (measured) current density $J(T)$ is used to discuss the pinning mechanism in pristine and proton-irradiated Ba(Fe$_{1-x}$Co$_x$)$_2$As$_2$.
However, identification of pinning mechanism using $J(T)$ is only empirical and lacks firm physical background.
So, although their conclusion and our conclusion on the pinning mechanism are similar,
we believe that our identification of pinning mechanism is more appropriate.

We can basically describe physical quantities for vortex system by means of collective creep theory both in the case of pristine and H$^{+}$-irradiated Ba(Fe$_{0.93}$Co$_{0.07}$)$_{2}$As$_{2}$.
It is noteworthy that the effect of H$^+$ irradiation can be summarized as $U_{00}$ enhancement, without replacing model function $U_0(T)/U_{00}$ and $J_0(T)/J_{00}$.
Namely, we can conclude that the effect of H$^{+}$ irradiation is enhancement of the collective pinning force by increasing weak-point-pinning centers without a drastic change of pinning mechanism.

Finally, we comment on the absolute value of $J$ instead of $J(T)/J(0)$.
$J$ is determined by the sum of weak-collective-pinning contribution $J^{\rm{wcp}}$ and strong-point-pinning contribution $J^{\rm{spp}}$.~\cite{PhysRevB.81.174517}
Instead, if we assign $J$ to $J^{\rm{spp}}$ in the absence of flux creep (so that we write $J$ as $J_c$ here),
we can estimate upper limit of strong-pinning center fraction in the crystal.
In the strong pinning theory~\cite{PhysRevB.66.024523}, critical current density is written as $J_c \approx 0.14\sqrt{n}\gamma [DF(T)]^{3/2}J_0$.
Here, $J_0 = c\phi_0/12\sqrt{3}\pi^2\xi_{ab}\lambda_{ab}^2$ is depairing current density, $\gamma = H_{c2}^{ab}/H_{c2}^{c}$ is anisotropy parameter, $n$, $D$ are density and diameter of pinning centers, respectively.
Assuming $D$ as several times of $\xi_{ab}$,
we can simplify $F(T) \approx \ln[1+D^2/8\xi^2(T)] \approx 1$.
Using pinning center volume $\nu \approx D^3/2$,
$n\nu = (J_c/J_0)^2(\sqrt{2}\times0.14\gamma)^{-2} \approx 0.05\%$ with $\xi_{ab} \sim 34$ \AA \;from $H_{c2}(0) \sim 280$ kOe~\cite{JPSJ.78.023702} and $\lambda_{ab} \sim 2000$ \AA.~\cite{PhysicaC.469.582}
This value is similar to the value reported in Na-doped CaFe$_2$As$_2$.~\cite{PhysRevB.84.094522}

In summary, we have studied the effect of proton irradiation up to $1.2\times 10^{16}$ cm$^{-2}$ in optimally Co-doped BaFe$_2$As$_2$ single crystals.
Critical current density under self-field is enhanced by a factor of 2.5 at 2 K.
Temperature dependence of critical current density and normalized flux relaxation rate is interpreted by collective creep theory.
With Maley's method, a glassy exponent $\mu \sim 1$ and variation of barrier height for flux creep $U_{00} \sim 41$ K to 93 K are directly determined.
To explain the value of $J$ after the creep from the model function of $J_c$,
$J_c$ is concluded to be controlled by both $\delta T_c-$ and $\delta l-$pinning. 
This model function is consistent with the result of generalized inversion scheme.
Proton irradiation effect is concluded as doubling of barrier height in the absence of flux creep.

\begin{acknowledgments}
This work was made as a part of the Research Project with Heavy Ions at
NIRS-HIMAC.
\end{acknowledgments}

%

\end{document}